\documentclass[final]{IEEEtran}

\usepackage{graphicx,epsfig,amsfonts,amsmath,amssymb,mathrsfs,color,enumerate,empheq,mathdots,empheq}
\usepackage[noadjust]{cite}
\usepackage{booktabs}
\usepackage{placeins}
\usepackage{comment}
\usepackage{stfloats}
\usepackage{floatrow}
\newfloatcommand{capbtabbox}{table}[][\FBwidth]
\usepackage{tikz}
\usepackage{placeins}
\usepackage[linesnumbered,vlined,ruled]{algorithm2e}
\usepackage{floatrow}
\newcommand{\subparagraph}{}
\usepackage{titlesec}
\usepackage[justification=centering]{caption}
\usepackage{subcaption}
\usepackage[usenames,dvipsnames]{pstricks}
\usepackage{hyperref}
\usepackage{dsfont}

\def\BibTeX{{\rm B\kern-.05em{\sc i\kern-.025em b}\kern-.08em
    T\kern-.1667em\lower.7ex\hbox{E}\kern-.125emX}}

\DeclareSymbolFont{grb}{OML}{cmm}{b}{it}
\DeclareMathSymbol{\zetab}{\mathord}{grb}{"10}
\DeclareMathSymbol{\etab}{\mathord}{grb}{"11}
\DeclareMathSymbol{\thetab}{\mathord}{grb}{"12}
\DeclareMathSymbol{\kappab}{\mathord}{grb}{"14}
\DeclareMathSymbol{\lambdab}{\mathord}{grb}{"15}
\DeclareMathSymbol{\mub}{\mathord}{grb}{"16}
\DeclareMathSymbol{\nub}{\mathord}{grb}{"17}
\DeclareMathSymbol{\rhob}{\mathord}{grb}{"1A}
\DeclareMathSymbol{\sigmab}{\mathord}{grb}{"1B}
\DeclareMathSymbol{\taub}{\mathord}{grb}{"1C}
\DeclareMathSymbol{\phib}{\mathord}{grb}{"1E}
\DeclareMathSymbol{\psib}{\mathord}{grb}{"20}
\DeclareMathSymbol{\omegab}{\mathord}{grb}{"21}
\DeclareMathSymbol{\epsilonb}{\mathord}{grb}{"22}
\DeclareMathSymbol{\varphib}{\mathord}{grb}{"27}

\DeclareFontFamily{U}{mathx}{\hyphenchar\font45}
\DeclareFontShape{U}{mathx}{m}{n}{
      <5> <6> <7> <8> <9> <10>
      <10.95> <12> <14.4> <17.28> <20.74> <24.88>
      mathx10
      }{}
\DeclareSymbolFont{mathx}{U}{mathx}{m}{n}
\DeclareFontSubstitution{U}{mathx}{m}{n}
\DeclareMathAccent{\widecheck}{0}{mathx}{"71}

\newfont{\Bd}{msbm10 at 12 truept}
\newfont{\Sc}{eusm10 at 12 truept}

\def\0{{\bf 0}}
\def\1{{\bf 1}}

\def\h{{\bf h}}

\def\r{{\bf r}}
\def\s{{\bf s}}
\def\t{{\bf t}}

\def\x{{\bf x}}

\def\E{{\bf E}}

\def\H{{\bf H}}
\def\I{{\bf I}}

\def\0{{\bf 0}}

\def\T{{\bf T}}

\def\bPhi{ {\boldsymbol \Phi} }

\def\mcF{{ \mathcal{F} }}

\def\mcJ{{ \mathcal{J} }}
\def\mcK{{ \mathcal{K} }}
\def\mcL{{ \mathcal{L} }}

\DeclareCaptionJustification{justified}{\justified}

\makeatletter
\def\ifundefined{\@ifundefined}

\makeatother

\begin{document}

      \title{ 
        Optimized Transmission Strategy for UAV-RIS 2.0 Assisted Communications Using Rate Splitting Multiple Access}
  
    \author{ Aamer Mohamed Huroon \textsuperscript{§}, Yu-Chih Huang \textsuperscript{†}, and Li-Chun Wang \textsuperscript{*},\\
\IEEEauthorblockA{\textit{\textsuperscript{{§}} 
            Department of Electrical Engineering and Computer Science International Graduate Program,\\ National Yang Ming Chiao Tung University } \\
    \textit{\textsuperscript{†} 
            Institute of Communications Engineering , National Yang Ming Chiao Tung University}\\
       \textit{  \textsuperscript{*} Department of Electrical and Computer Engineering, National Yang Ming Chiao Tung University } \\
       \textsuperscript{§} aamer.ee08@nycu.edu.tw,
    \textsuperscript{†} jerryhuang@nycu.edu.tw,
   \textsuperscript{*} wang@nycu.edu.tw
       }
    \thanks{ This work has been partially funded by the National Science and Technology Council under the Grants MOST 110-2221-E-A49-039-MY3, and MOST 111-2221-E-A49-071-MY3, and NSTC 111-2634-F-A49-010, and NSTC 111-3114-E-A49-001, Taiwan. This work was also financially supported by the Center for Open Intelligent Connectivity from The Featured Areas Research Center Program within the framework of the Higher Education Sprout Project by the Ministry of Education (MOE) in Taiwan. This work was supported by the Higher Education Sprout Project of the National Yang Ming Chiao Tung University and Ministry of Education (MOE), Taiwan.}} 
    \maketitle

     \begin{abstract}
    In this paper, we study the transmission strategy of a ground-based beyond diagonal reconfigurable intelligent surface (BD-RIS), a.k.a  RIS 2.0, in a network where multiple unmanned aerial vehicles (UAVs) simultaneously transmit signals to the respective groups of users. It is assumed that each group is assigned subcarriers orthogonal to those assigned to other groups and rate splitting multiple access (RSMA) is adopted within each group. A corresponding mixed integer nonlinear programming problem (MINLP) is formulated, which aims to jointly optimize 1) allocation of BD-RIS elements to groups, 2) BD-RIS phase rotations, 3) rate allocation in RSMA, and 4) precoders. To solve the problem, we propose using generalized benders decomposition (GBD) augmented with a manifold-based algorithm. GBD splits the MINLP problem into two sub-problems, namely the primal and the relaxed master problem, which are solved alternately and iteratively. In the primal problem, we apply block coordinate descent (BCD) to manage the coupling of variables effectively. Moreover, we recognize the manifold structure in the phase rotation constraint of BD-RIS, enabling the Riemannian conjugate gradient (RCG).
    Simulation results demonstrate the effectiveness of the proposed approach in maximizing spectral efficiency. 

    \end{abstract}


    \section{Introduction}\label{sec_into}

    Using multiple unmanned aerial vehicles (UAVs) as base stations in 6G wireless communications offers several advantages, such as rapid deployment, flexibility, mobility, and expanded coverage. However, to achieve advanced communication features like high data rates, low latency, and connectivity for numerous devices, UAVs must be integrated with other technologies rather than functioning alone \cite{ge2023intelligent,huroon2022generalized,huroon2023generalized }.

    Another emerging technology is reconfigurable intelligent surfaces (RIS) technology which holds great promise for improving wireless communication systems. RIS can manipulate the propagation of electromagnetic fields, offering potential benefits in enhancing wireless communication capabilities \cite{wu2019intelligent, cao2021reconfigurable}.
    Recent advancements in reconfigurable intelligent surfaces (RIS) technology have led to the development of beyond-diagonal reconfigurable intelligent surfaces (BD-RIS), a.k.a. RIS 2.0. Unlike conventional RIS, which restricts the phase rotation matrices to be diagonal, BD-RIS explores more sophisticated techniques to unleash more advantages by enabling more possibilities for phase rotation matrices \cite{li2022beyond,shen2021modeling}.  
    
    While integrating unmanned aerial vehicles (UAVs) with BD-RIS may address coverage problems caused by an increasing number of users, it would also introduce a new challenge of multi-user interference. To tackle this issue, a solution has been proposed \cite{soleymani2023optimization}, involving the integration of rate splitting multiple access (RSMA) \cite{mao2022rate} into the system, aiming to handle the problem of multi-user interference effectively. 
    
    To harness the potential of the considered integration system effectively, we introduce a joint optimization approach aimed at maximizing the overall achievable sum rate. We are examining a system in which several aerial UAVs function as base stations, attempting to serve clusters of ground users with the help of a ground-based BD-RIS, which is partitioned into BD-RIS cells and then assigned exclusively to a cluster. Moreover, each cluster is allocated subcarriers that are orthogonal to those allocated for other clusters, while RSMA is adopted with each cluster to manage inter-cluster interference. This gives rise a new optimization problem that involves finding the optimal configurations for the following variables: 1) BD-RIS phase rotation matrix, 2) BD-RIS cell allocation, 3) precoder optimization, and 4) common rate allocation in RSMA. We note that compared to our prior work \cite{huroon2023generalized} which investigates a similar optimization problem for the same network model with the help of a conventional RIS and conventional non-orthogonal multiple access, the present work is a generalization that further incorporates more advanced BD-RIS and RSMA. 
    
    The newly formulated problem is complex, combining discrete and continuous variables, making it a mixed integer nonlinear programming problem (MINLP). To address this challenge, we put forward a solution using generalized Benders decomposition (GBD) augmented with a manifold-based algorithm. To demonstrate the effectiveness of the proposed system model, extensive simulations are carried out. The results demonstrate that the integration of UAV-BD-RIS with the RSMA approach significantly outperforms conventional system models, including conventional RIS and the NOMA scheme. The outcomes highlight the significance of BD-RIS and RSMA technologies as essential drivers for enhancing spectral efficiency while establishing sustainable UAV base station networks.

    \subsection{Paper Organization}

    The subsequent sections of the paper are structured as follows: Section~\ref{sec_bac} presents a discussion on the background knowledge. Section~\ref{sec_meth} outlines the formulation and solution of the proposed problem. In Section~\ref{sec_sim}, simulation results are presented. Finally, Section~\ref{sec_con} offers concluding remarks.

    \subsection{Notation}
The operation $(\cdot)^{H}$ represents the conjugate transpose, of a matrix or vector. The symbol $|\cdot|$ refers to the absolute value of a complex number. Boldface characters represent either column vectors or matrices. The notation $\mathbb{C}^{x \times y}$ denotes the set of all $x \times y$ complex-valued matrices. Table I offers a summary of the essential notations. For any natural number $L$, we represent the set of integers (indices) as $[L]$.



    \begin{figure}
             \includegraphics[width=0.7\linewidth]{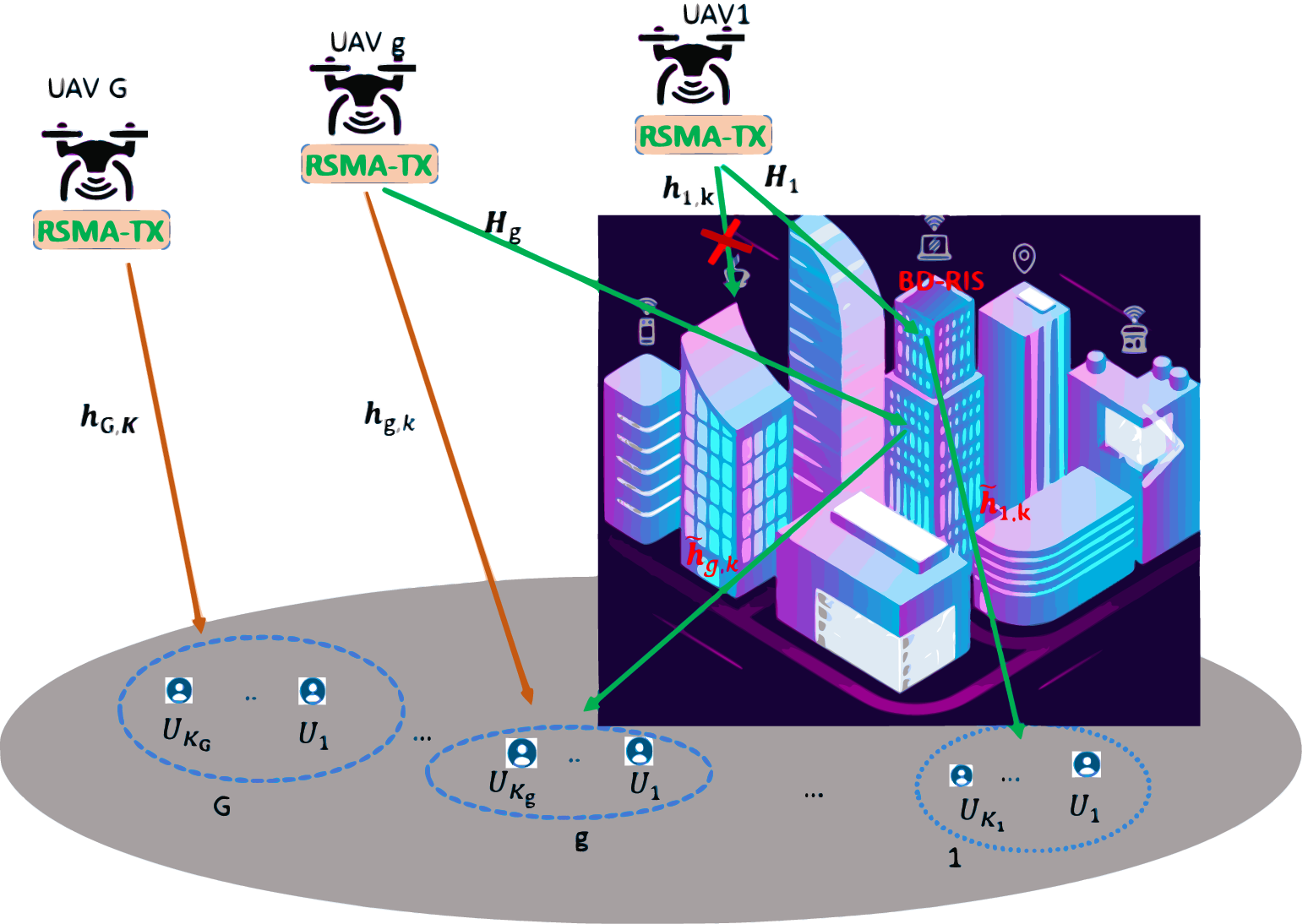}
                    \caption{The UAV- BD-RIS assisted system.}
                    \label{fig:system_model}
    \end{figure}

    \section{Background }\label{sec_bac}
    This section presents the system model and channel model in Sections~\ref{subsec_system_model} and \ref{subsec_ch_model}, respectively. 
    \subsection{System Model}\label{subsec_system_model}
    The considered system is a multi-user downlink aerial-terrestrial communication setup with support from a BD-RIS. The system comprises multiple UAVs that serve distinct groups of users. Each UAV is set responsible for serving a specific group, and the BD-RIS with a cell-wise fully connected (CW-FC) architecture, is installed on a building facade, consisting of passive reflecting cells that can be adjusted individually using low-cost devices. 

    In BD-RIS, the reflecting cells are organized into several clusters, and each cluster supports a subset of UAVs. The structure of the BD-RIS matrix $\bPhi$ is determined by the circuit topology of the $2L$-port reconfigurable impedance network. $\bPhi=\{\bPhi_g:~g\in[G]\}$, and meet the constraint,
    $        
    \boldsymbol{\Phi}_{g}^H \boldsymbol{\Phi}_{g}=\mathbf{I}_L.
    $
     
    To elaborate further, there are a total of $G$ UAVs, each serving a distinct group of users denoted as $\mathcal{K}_g$, and each UAV is equipped with $N$ antennas. The UAVs are divided into $F$ clusters, indexed as $\mathcal{F}_1, \ldots, \mathcal{F}_F$, with each cluster having $L/F$ reflecting cells. The BD-RIS operates passively, with only the controller and configuration circuits requiring power.

    The assignment of UAVs to BD-RIS clusters is denoted by $u_g$, where $u_g = f$ indicates that UAV $g$ is assisted by RIS cluster $f$, and $u_g = 0$ means that UAV $g$ is not assisted by any RIS cluster. We use $f(u_g)$ as an indicator function to determine if UAV group $g$ is assisted by the BD-RIS or not. Consequently, the total number of active BD-RIS clusters is $F = \sum_{g=1}^G f(u_g)$.

    The system's bandwidth is divided into $C$ orthogonal sub-carriers, and each UAV occupies sub-carriers that do not interfere with others' sub-carriers. The $C$ sub-carriers are further divided into two parts: $\omega_{1} C$ and $\omega_{2} C$, with $\omega_{1} + \omega_{2} = 1$. The $\omega_{1} C$ part is assigned for RIS-aided transmissions, allocated to UAV groups with $f(u_g) = 1$, while the remaining $\omega_{2} C$ is used for UAV groups without BD-RIS assistance, i.e., $f(u_g) = 0$.

    For UAVs ($g\in[G]$) assisted by BD-RIS ($f(u_g)=1$), the system considers channel vectors $\H_{g}\in\mathbb{C}^{L/F \times N}$ and $\tilde{\h}_{g,k}\in\mathbb{C}^{1 \times L/F}$, representing the channels between UAV $g$ and the BD-RIS cells in the assigned cluster $u_g$, and between the BD-RIS cells in cluster $u_g$ and the user $k\in\mcK_g$, respectively. Additionally, $\h_{g,k}\in\mathbb{C}^{1\times N}$ represents the direct channel between UAV $g$ and user $k\in\mcK_g$.

    The communication process involves a two-phase transmission protocol \cite{cao2021reconfigurable}, comprising a negotiation phase and a communication phase.
     During the negotiation phase, it is assumed that all channel estimation and synchronization are done perfectly \cite{cao2021reconfigurable, wu2019intelligent}.
    On the other hand, the communication phase is used for actual data transmission from the UAVs, where the UAV $g\in[G]$ sends   
    \[
    \s_g = ( \t_{g}^c x_{g}^c  + \sum_{k \in \mathcal{K}_g}\t_{g,k}^p x_{g,k}^p)  =
        \T_g \x_g,
    \]
    where $\t_{g}^c \in \mathbb{C}^{N \times 1}$ , $ x_{g}^c  \in \mathbb{C}$, denotes the precoder vector and transmit symbol for the common part respectively. $\t_{g,k}^p \in \mathbb{C}^{N \times 1}$ , $x_{g,k}^p \in \mathbb{C}$ denotes the precoder vector and transmit symbol for the private part respectively.
    $\T_g = [\t_g^c, \t_{g, 1}^p, \ldots, \t_{g, K}^p] \in \mathbb{C}^{ N \times(K+1)}$, is the precoder matrix
 and $\x_g \in \mathbb{C}^{(K+1) \times 1}$ represents the transmitted signals, distributed as $\x_g \sim \mathcal{C N}\left(\mathbf{0}, \I_{(K+1)}\right)$.
    
     The expression for the received signal at user $k$ from the $g$-th UAV can be represented as follows:
        
\begin{equation}\label{eq_re_s}
    \begin{split}
       &y_{g,k}  =
              \Big(\h_{g,k} +
             {\tilde{\h}}_{g,k}  \bPhi_g\H_{g} \Big) \s_{g}
              +\; w_{g,k}\\
      & =  \underbrace{
            \Big(\h_{g,k} +
             {\tilde{\h}}_{g,k} \bPhi_g\H_{g} \Big) \t_{g}^c x_{g}^c }_{common\; part}
+\; \underbrace{
         \Big(\h_{g,k} +
             {\tilde{\h}}_{g,k}  \bPhi_g\H_{g} \Big)\t_{g,k}^p x_{g,k}^p, }_{private\; part} \\
     &       +\;  
           \underbrace{ \sum_{\substack{j\in \mathcal{K}_g\setminus k}}\Big(\h_{g,k} +
             {\tilde{\h}}_{g,k}  \bPhi_g\H_{g} \Big)\t_{g,j}^p x_{g,j}^p  
        }_{interference} + \;\underbrace{ w_{g,k}}_{AWGN},
    \end{split}
\end{equation}
where the additive white Gaussian noise can be represented as $\omega_{g,k} \sim \mathcal{C N}\left(0, \sigma^{2} \right)$. The scattering matrix $\bPhi_g$ comprises phase shift values $f_1,\ldots,f_{L/F}\in\mcF_{u_g}$ for BD-RIS cells within BD-RIS cluster $u_g$, where $u_g>0$. If $u_g=0$, $\bPhi_g$ is set to $\mathbf{0}$. Some channel gains $\h_{g,k}$ may be equal to $0$, indicating that the user is within a coverage hole.


    This paper aims to optimize the assignment of $u_g$ and $\bPhi_g$ for $g\in[G]$, which includes the allocation of BD-RIS cells represented by $u_g$ and the corresponding phase rotation matrix $\bPhi_g$, with the primary objective of maximizing the system's achievable sum rate. It is important to mention that a specific scenario of the problem has been explored in prior research, where $K_g=1$ for all $g\in[G]$, as discussed in \cite{cao2021reconfigurable}. In this paper, we will investigate both this special case and the more general scenario with arbitrary values of $K_g$. As mentioned earlier, $F$ groups of UAVs will receive assistance from the BD-RIS, while the remaining $G-F$ UAV groups will solely rely on the direct link for communication. Consequently, each UAV group $g$ with a binary function $f(u_g)=1$ will utilize a fraction of $\omega_1/F$ of the available bandwidth. Conversely, each UAV group $g$ with $f(u_g)=0$ will utilize $\omega_2/(G-F)$ fraction of bandwidth.

    \subsection{ Channel Model}\label{subsec_ch_model}

  As previously indicated, \eqref{eq_re_s} deals with multiple channels that exhibit diverse characteristics. These channels encompass the transmission from the UAV to the RIS, the transmission from the RIS to the users, and the direct transmission from the UAV to the users. Drawing on earlier studies in the realm of RIS \cite{cao2021reconfigurable, 9110889}, we employ a Rician fading channel model for all these channels. This choice of model proves to be more comprehensive since the presence of the UAV facilitates a line-of-sight (LOS) link.

    \section{Problem Formulation and Proposed Solution }\label{sec_meth}

    In this section, our focus is on the RSMA acting as an intermediary multiple access scheme between two other schemes, treating interference as noise and full decoding. We begin by presenting the optimization problem in Section~\ref{subsec_problem}. Next, the proposed solution is detailed in Sections~\ref{subsec_proposed_GBD}, respectively.

    \subsection{The Problem}\label{subsec_problem}

    Based on the expression in \eqref{eq_re_s}, the signal-to-interference-plus-noise ratio (SINR) for the UAV groups assisted by the RIS (i.e., $f(u_g)=1$) is given by: 
   \begin{align}
      \gamma_{g,k}^c &= \frac{|(\h_{g,k} +      {\tilde{\h}}_g,k{}  \bPhi_g\H_{g})\t_{g}^c |^2}
      {\sum_{k \in \mathcal{K}_g}  |(\h_{g,k} +      {\tilde{\h}}_{g,k}  \bPhi_g\H_{g})\t_{g,j}^p|^2 +  \sigma_k^2},\label{eq_sinr_reflect} \\
      \gamma_{g,k}^p &= \frac{|(\h_{g,k} +      {\tilde{\h}}_{g,k}  \bPhi_g\H_{g})\t_{g,k}^p |^2}
      {\sum_{k \in \mathcal{K}_{g\backslash k}} |(\h_{g,k} +      {\tilde{\h}}_{g,k}  \bPhi_g\H_{g})\t_{g,j}^p|^2 +  \sigma_k^2}.\label{eq_sinr_reflect}
    \end{align}

    Likewise, according to the expression in \eqref{eq_re_s}, the signal-to-interference-plus-noise ratio (SINR) for UAV groups that do not receive assistance from the RIS (i.e., $f(u_g)=0$) is given by: $\forall k \in \mathcal{K} \text {, }$

   \begin{align}
     \Bar{\gamma}_{g,k}^c &=  \frac{|\h_{g,k}\t_{g}^c|^2}{\sum_{k \in \mathcal{K}_g} |\h_{g,k} \t_{g,j}^p|^2 + \sigma_k^2},\\
      \Bar{\gamma}_{g,k}^p &=  \frac{|\h_{g,k}\t_{g,k}^p|^2}{\sum_{k \in \mathcal{K}_{g\backslash k}}  |\h_{g,k} \t_{g,j}^p|^2 + \sigma_k^2}.
    \end{align}

    Consider sets $\mathcal{U}=\{u_g:~g\in [G]\}$ and $\bPhi=\{\bPhi_g:~g\in[G]\}$.
    As a reminder, for UAV groups where $f(u_g)=0$, the corresponding matrix $\bPhi_g$ is $\mathbf{0}$. 
    Taking everything into account, the achievable sum rate for common and private parts respectively is given by: 
  
    \begin{align}
    &R_{g, k}^c =\log _2\Big(1 + \frac{|(\h_{g,k} + \tilde{\h}_{g,k}  \bPhi_g\H_{g})\t_{g}^c |^2}{\sum_{k \in \mathcal{K}_g}  |(\h_{g,k}+ \tilde{\h}_{g,k}^H\bPhi_g\H_{g})\t_{g,j}^p|^2 + \sigma_k^2}\Big),\\
   &R_{g, k}^p = \;\log _2\Big(1+ \frac{|(\h_{g,k} +\tilde{\h}_{g,k}  \bPhi_g\H_{g})\t_{g,k}^p |^2}
      {\sum_{k \in \mathcal{K}_{g\backslash k}}  |(\h_{g,k}^H + \tilde{\h}_{g,k}  \bPhi_g\H_{g})\t_{g,j}^p|^2 + \sigma_k^2} \Big).
   \end{align}

    Similarly, the achievable sum rate for UAVs that do not receive assistance from the BD-RIS through the direct path is expressed as follows for both the common and private parts:
     \begin{align}
\Bar{R}_{g, k}^c &=\log _2\left(1+
   \frac{|\h_{g,k}\t_{g}^c|^2}{\sum_{k \in \mathcal{K}_g}  |\h_{g,k} \t_{g,j}^p|^2 + \sigma_k^2}
    \right), \\
  \Bar{R}_{g, k}^p &= \log _2\left(1+ 
   \frac{|\h_{g,k}\t_{g,k}^p|^2}{\sum_{k \in \mathcal{K}_{g\backslash k}}  |\h_{g,k} \t_{g,j}^p|^2 + \sigma_k^2} \right).
   \end{align}

    It is important to note that the common signal is fully decoded by all users, and its rate should not surpass the channel capacity. This can be mathematically expressed as:

    $\sum_{i=1}^{K_g} r_{g, i} \leq R_{g, k}^c, \forall k \in \mathcal{K}$.
    Here, ${r}_g=\left[r_{g, 1}, \cdots, r_{g, K}\right]^T \in \mathbb{C}^K$ represents the common rate allocation vector.

    The overall achievable sum rate for all common and private parts of UAVs assisted by BD-RIS  and UAVs not assisted by BD-RIS respectively it can be given as
    \begin{align}
    R_{\text {overall }}&(\mathcal{U},\bPhi_g,\T,\r_g) =  \sum_{g=1}^{G}\sum_{k=1}^{K_g} \frac{\omega_1 C}{F} f(u_{g})  
    \Big(  R_{g,k}^c + R_{g,k}^p   \Big) \nonumber \\   
 & + \frac{\omega_2 C}{G-F}  (1-f(u_{g})) 
  \Big(\Bar{R}_{g,k}^c + \Bar{R}_{g,k}^c \Big).
    \label{eq_OVE_RATE}
    \end{align}

   
    Having expressed \eqref{eq_OVE_RATE}, our objective is now to create an optimization problem that maximizes $R_{\text{overall}}$ by jointly designing RIS cell allocation $\mathcal{U}$, precoder matrix optimization $\T_g$, common rate allocation $\r_g$, and BD-RIS phase rotation $\bPhi_g$. The specific formulation of the problem can be stated as follows:
      \begin{equation}
    \begin{split}
    & \max_{ \mathcal{U} , \bPhi_g,\T_g,\r_g} \;~R_\text{overall}(\mathcal{U}, \bPhi_g,\T,\r_g), \\
    \text{s.t.} &\;
        \mathsf{C}_1:u_g \in [F]\cup {0},\forall g \in [G],\\
         &\; \mathsf{C}_2: u_g \neq u_{g'},\forall g,g'\in[G],
         \text{ with } f(g)=f(g')=1,\\
          &\; \mathsf{C}_3: \sum_{g=1}^{G} f(u_g) \leq F,  \forall g \in [G],\\
    &\; \mathsf{C}_4: \sum_{j=1}^{K_g} r_{g, j}^c \leq R_{g, k}^c,\; r_{g, k}^c \geq 0, \forall k \in \mathcal{K}_g, \forall g \in [G],\\
     &\; \mathsf{C}_5: r_{g, k}^c + R_{g, k}^p \geq R^{\min }, \quad \forall k \in \mathcal{K}_g,  \forall g \in [G], \\ 
         &\; \mathsf{C}_6: \sum_{k=0}^{K_g}\left\|\mathrm{\t}_{k}^{g} \right\|^2 \leq P_{\mathrm{UAV}}^{\max },\;\forall g \in [G], \\
                             &\; \mathsf{C}_7: \boldsymbol{\Phi}_{g,f_l}^H \boldsymbol{\Phi}_{g,f_l}=\mathbf{I}_L,   \forall g \in [G],
                             \text{ s.t. $f(u_g)=1$},\forall f_l \in \mathcal{F}_{u_g},\label{eq_ORG_P}
    \end{split}
    \end{equation}
    where the formulated problem includes the following constraints:
$\mathsf{C}_1$-$\mathsf{C}_3$ handle the assignment of users to the RIS clusters, 
$\mathsf{C}_4$ ensures that all users can accurately decode the common signal,
$\mathsf{C}_5$ guarantees that all users meet the required minimum rate,
$\mathsf{C}_6$ represents the power constraint at the base station, and finally
$\mathsf{C}_7$ represents the most general constraint related to the BD-RIS phase shift.

   \subsection{Problem Solution}\label{subsec_proposed_GBD}
    Equation \eqref{eq_ORG_P} corresponds to an optimization problem categorized as MINLP. To tackle this complex issue, we will employ the GBD algorithm, which divides it into two sub-problems: the primal problem and the relaxed master problem.

   \subsubsection{Primal problem: Solving $\bPhi_g$, $\T_g$ and $\r_g$ with fixed $\mathcal{U}$}

    At iteration $\ell$, with a fixed $\mathcal{U}=\mathcal{U}^{(\ell-1)}=\{u_1^{(\ell-1)},\ldots,u_F^{(\ell-1)}\}$, \eqref{eq_ORG_P} becomes

      \begin{equation}
    \begin{split}
      \max_{\bPhi_g,\T_g,\r_g} \;&\;R_{\text {overall }}(\mathcal{U}^{\ell-1},\bPhi_g,\T_g,\r_g )
            \; \text{s.t.} \;   \mathsf{C}_4- \mathsf{C}_7.
    \label{eq_APL_RCG}
    \end{split}
     \end{equation}
    In order to further decompose the primal problem presented in \eqref{eq_APL_RCG}, we encounter the challenge of coupling three variables. To address this complexity, we propose employing the block coordinate descent  (BCD) method. The BCD method allows us to break down the problem into manageable blocks, facilitating the solution process. Then problem \eqref{eq_APL_RCG}, becomes.


  \begin{equation}
    \begin{split}
        \max_{\bPhi_g} &\;R_{\text {overall }}(\mathcal{U}^{\ell-1},\bPhi_g,\hat{\T_g},\hat{\r_g} )\;
             \text{s.t.} \;    \mathsf{C}_7.
    \label{eq_MAX_aPH}
    \end{split}
     \end{equation}
    
    We note that the objective function in \eqref{eq_MAX_aPH} is continuous and differentiable. Moreover, the constraint sets of \eqref{eq_MAX_aPH} form a complex circle manifold. Hence, the problem becomes optimization on a manifold and can be solved by algorithms such as the Riemannian conjugate gradient (RCG) methods \cite{boumal2014manopt, guo2020weighted}.
  Once the matrix $\bPhi_g$ is kept constant, the optimization of $\mathbf{T_g}$ becomes a simplified form of the weighted sum rate maximization problem in a conventional multi-user multiple-input single-output system. This problem has been extensively studied in the literature, and one well-known approach to finding the stationary solution is through the minimum mean square error (MMSE) algorithm \cite{shi2011iteratively,liu2020simple}, which utilizes the following iterative updating rule: 
     \begin{equation}
    \begin{split}
        \max_{\T_g,r_g} &\;R_{\text {overall }}(\mathcal{U}^{\ell-1},\hat{\bPhi_g},\T_g,\r_g ).\;
             \text{s.t.} \;   \mathsf{C}_4- \mathsf{C}_6.
    \label{eq_MAX_bPH}
    \end{split}
     \end{equation}
     
    The MMSE algorithm can effectively address this problem using the iterative updating rule. The common rate allocation for the rate splitting vector $\mathbf{r}_g$ is solved concurrently with the precoder matrix. This involves finding optimal solutions for both the rate allocation and the precoder matrix to achieve the desired outcomes.

  By utilizing BCD with equations \eqref{eq_MAX_aPH}, and \eqref{eq_MAX_bPH}, we solve the primal problem. If feasible, this problem provides a lower bound on the original problem \eqref{eq_ORG_P}, denoted as $\mathsf{LB}^{(\ell)}$, along with the corresponding phase rotation matrix $\bPhi^{(\ell)}=\{\bPhi_g^{(\ell)}:g\in[G]\}$. The iteration index $\ell$ is included in the feasible set $\mcJ$, and the optimal dual variable ${\boldsymbol\mu}^{(\ell)}$ is retained for future use.
    If the primal problem is infeasible, we add $\ell$ to the infeasible set $\bar{\mcJ}$ and formulate a feasibility check problem. We keep the Lagrangian multiplier ${\boldsymbol\lambda}^{(\ell)}$ associated with this feasibility problem for later use.

   \subsubsection{Master problem: Solving $\mathcal{U}$ with $\bPhi_g$, $\T$ and $\r_g$}
     During iteration $\ell$, we keep $\bPhi^{(\ell)}=\{\bPhi_g^{(\ell)}:g\in[G]\}$ fixed and proceed to find the optimal $\mathcal{U}$ by solving the following form of the original problem \eqref{eq_ORG_P}:
    
     \begin{equation}
    \begin{split}
      \max_{\mathcal{U}} &\;R_{\text {overall }}(\mathcal{U},\bPhi_g^{\ell-1},\T_g^{\ell-1},\r_g^{\ell-1} )\;
       \text{s.t.} \; \mathsf{C}_1-\mathsf{C}_3 .
    \label{eq_MAX_ELE}
     \end{split}
     \end{equation}

    The GBD algorithm uses optimality cuts to update a feasible primal problem and feasibility cuts to update an infeasible primal problem. The Lagrangian function for feasible and infeasible primal problems respectively written as follows:

   \begin{equation}
     \begin{split}
           \boldsymbol{\mcL}(\T_g, \r_g,{{\bPhi_g}},\mathcal{U},\boldsymbol{{\mu}}) & =
                \;R_{\text {overall }}(\mathcal{U},\bPhi_g,\T_g,\r_g )\\
                  & + \; \boldsymbol{{\mu}} \E(\mathcal{U},\T_g, \r_c,{{\bPhi_g}}), \;\boldsymbol{{\mu}}\succeq \mathbf{0}.
    \end{split}
    \label{eq_DUAL_ITN}
    \end{equation}
    \begin{equation}
     \begin{split}
           &\Bar{ \boldsymbol{\mcL}}(\T_g, \r_g,{{\bPhi_g}},\mathcal{U},\boldsymbol{{\lambda}})  =
                  \boldsymbol{{\lambda}} \E(\mathcal{U},\T_g, \r_g,{{\bPhi_g}}),\;\boldsymbol{{\lambda}}\succeq\mathbf{0}.
    \end{split}
    \label{eq_DUAL_ITN}
    \end{equation}

    Where $\E(\mathcal{U}, \T_g, \r_g, \bPhi_g)$, represents the constraint functions in vector form. The Lagrangian multiplier for feasible and infeasible cases respectively denotes $\boldsymbol{{\mu}}$, and $\boldsymbol{{\lambda}}$. Finally, the relaxed master problem is written as follows:

    \begin{equation}
    \begin{split}
            \max_{\boldsymbol{\mathcal{U}},\eta}& \text{ } \eta, \\
                \text{s.t.}
                       \;  \;&\eta \leq \boldsymbol{\mcL}(\boldsymbol{{{\bPhi_g}}}^{(j)},\T_g^{(j)},\r_g^{(j)},\mathcal{U},\boldsymbol{{\mu}}^{(j)}),\forall j\in \mathcal{J}, \\
                            \;& 0 \leq  \Bar{ \boldsymbol{\mcL}}(\boldsymbol{{{\bPhi_g}}}^{(j)},\T_g^{(j)},\r_g^{(j)},\mathcal{U},\boldsymbol{{\lambda}}^{(j)}), \forall j \in \bar{\mathcal{J}}.
    \end{split}
     \label{eq_final_max_ITN}
    \end{equation}
At each iteration, the solution $\mathcal{U}^{(\ell)}$ is updated until it reaches the optimal solution, which provides an upper bound $\mathsf{UB}^{(\ell)}$, for the original problem \eqref{eq_ORG_P}.



    \section{Simulation Results}\label{sec_sim}

    In this section, we evaluate the performance of our proposed technique, through extensive simulation experiments. In Section~\ref{subsec_setting}, we describe the specific details of the simulation setup, and in Section~\ref{subsec_results}, we present the simulation results.

    \subsection{Simulation Setting}\label{subsec_setting}
    
     We work within a three-dimensional space with dimensions of $100m \times 100m \times 300m$. 
     The initial positions of UAV 1 and user 1 are fixed. For UAVs $2-8$, their $y$ and $z$ coordinates are common at $80m$ and $250m$, respectively, but each has a unique $x$ coordinate ranging from $10m$ to $100m$ in increments of $10m$. Similarly, all other users have the same $y$ and $z$ coordinates as user 1, i.e., $30m$ and $1m$ respectively, while their $x$ coordinates range from $15m$ to $85m$.
     
     
    
    \begin{table}[!th]
     \label{tab:my_label}
    \footnotesize
    \begin{center}
    \caption{Simulation Setting}
     \begin{tabular}{|p{2.1 cm}|p{2.1cm}|p{1.8 cm}|p{1.2cm}|  }
    \hline
        Notation  &  Definition &  Notation  &  Definition  \\
    \hline 
        BD-RIS location& $(100,75,120)m$&  $L$ & 512 \\
    \hline
        UAV1 location& (20,80,250)m&  $G$ & 8\\
    \hline
     User1 location& $(10,30,1)m$&  $N$ & 4\\
    \hline
     GBD max itr& $50$&   $  \sigma^{2} $ & $ -94 \mathrm{dBm} $\\
    \hline
       $F$& [1\;2\;4\;8] & $ \rho_g^{2} $ & $10 \mathrm{~mW} $ \\
    \hline 
        Frame length  & $ 1 \mathrm{~ms} $ &  Bandwidth  & 10 $ \mathrm{MHz} $ \\
    \hline
      Users  location   & ($x$,30,1) $\mathrm{m}$  &  Frequency  & $ 5 \mathrm{GHz} $ \\
    \hline 
        UAVs  location   & ($x$,80,250) $ \mathrm{m} $ & Antenna-gain  & 5 $ \mathrm{dBi} $ \\
    \hline
    BCD max itr  & $ 80$ &  $ \left(\omega_{1}, \omega_{2}\right) $ & $0.6,0.4$ \\
    \hline 
    \end{tabular}
    \end{center}
    \end{table}

    \subsection{Simulation Results}\label{subsec_results}
 
\begin{figure}[ht]
	\includegraphics[width=0.7\linewidth]{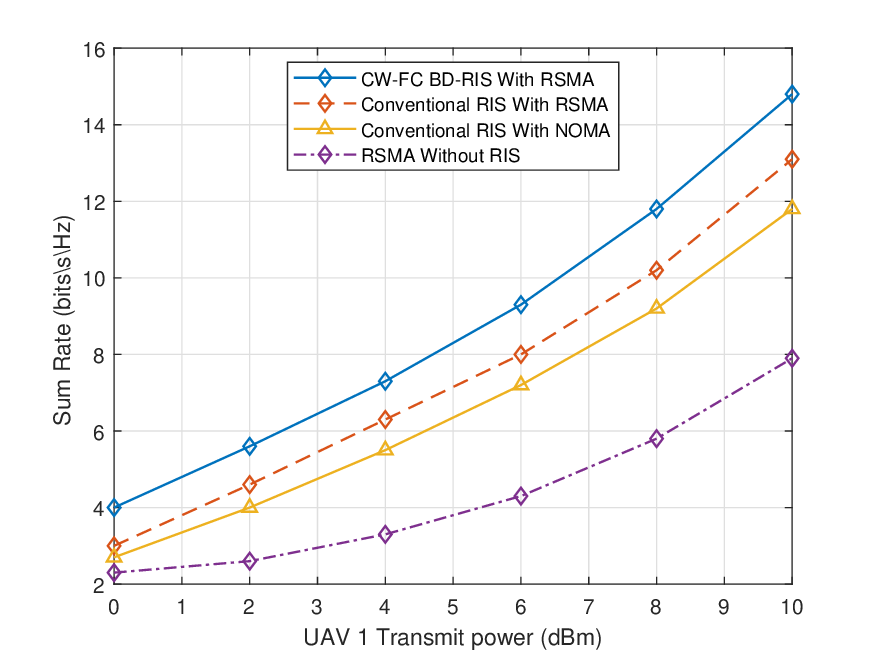}
    \caption{Achievable sum rate versus UAV transmit power. }
    \label{fig:TVU_SIMF}
    \end{figure}

      Fig.~\ref{fig:TVU_SIMF} presents the sum rate as a function of the transmit power of UAVs. The figure showcases the performance of various schemes, the proposed RSMA with CW-FC BD-RIS architectures, conventional RIS with RSMA, and conventional RIS with NOMA. The results indicate that the RSMA scheme with BD-RIS architecture outperforms the schemes with conventional RIS. It is observed that conventional RIS with RSMA outperforms conventional RIS with NOMA, this superiority is expected since NOMA can be deemed as a special case of RSMA. Moreover, RSMA without RIS performs the worst, highlighting the effectiveness of RIS in improving spectral efficiency.

\begin{figure}[ht]
	\includegraphics[width=0.7\linewidth]{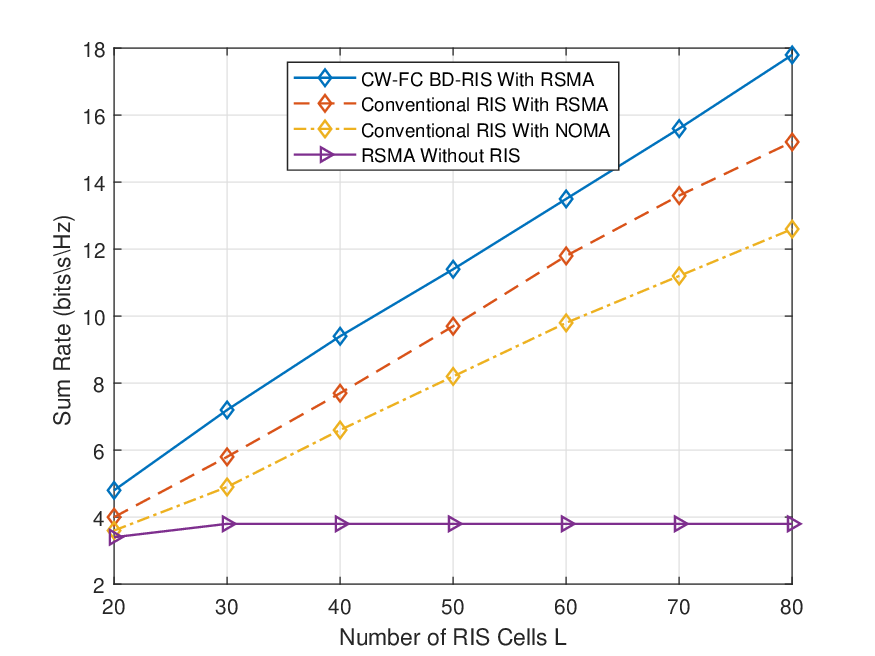}
    \caption{Achievable sum rate as a function of BD-RIS cells.}
    \label{fig:SIMS}
    \end{figure}

    Fig. ~\ref{fig:SIMS} illustrates the sum rate as a function of the number of BD-RIS cells. The results indicate that the sum rate increases with an increasing the number of BD-RIS cells for both RSMA and NOMA schemes. Notably, the proposed RSMA scheme outperforms the NOMA scheme in terms of achieving a higher sum rate across all values of BD-RIS cells. Additionally, the comparison between BD-RIS and conventional RIS shows that using BD-RIS can lead to better rates. Overall, these findings demonstrate the effectiveness of RSMA and BD-RIS in exploiting a large achievable sum rate, with an increasing number of BD-RIS cells.


    \section{Conclusion}\label{sec_con}
    This paper considered a UAV-BD-RIS assisted multi-user communication system, where several UAVs serve their respective cluster of users with the RSMA technique for interference management. A new optimization problem was formulated that aims at maximizing the sum rate over the choice of 1) BD-RIS matrix configuration, 2) BD-RIS cell allocation, 3) optimal precoder vector optimization, and 4) common rate allocation in RSMA. Recognizing the manifold structure in the phase rotation constraint, we propopsed a novel GBD algorithm that involves BCD and RCG when solving the primal sub-problem.  
    Simulation results were provided to demonstrate that the proposed integrated system can significantly enhance performance and achieve higher spectral efficiency. Future research may study the convergence of the proposed algorithm in UAV-BD-RIS assisted system.
 
    \bibliographystyle{IEEEtran}
    \bibliography{Ref.bib}
    \end{document}